\documentclass[twocolumn,showpacs,preprintnumbers,amsmath,amssymb]{revtex4}
\usepackage{graphicx}% Include figure files
\usepackage{dcolumn}% Align table columns on decimal point
\usepackage{bm}% bold math

\begin{document}

\preprint{APS/123-QED}

\title{Adiabatic Pair Creation in Heavy Ion and Laser Fields}% Force line breaks with \\

\author{Pickl, P. }

\email{pickl@mathematik.uni-muenchen.de}

\affiliation{Institut f\"ur Theoretische Physik der Universit\"at Wien\\
Boltzmanngasse 5\\
1050 Wien\\Austria% with \\
}%

\author{D\"urr, D.}
\email{duerr@mathematik.uni-muenchen.de}
\affiliation{ Mathematisches Institut der Universit\"at M\"unchen\\
Theresienstr. 39\\
80333 M\"unchen\\Germany% with \\
}%

\date{\today}% It is always \today, today,

\begin{abstract}
The  planned generation of lasers  and heavy ion colliders renews
the hope to see electron-positron pair creation in strong
classical fields. This old prediction is  usually referred to as
spontaneous pair creation. We observe that both---heavy ion
collisions and pair creation in strong laser fields---are
instances of the theory of adiabatic pair creation. We shall
present the theory, thereby correcting earlier results. We give
the momentum distribution of created pairs in overcritical fields.
We discuss carefully the proposed experimental verifications and
conclude that pure laser based experiments are highly
questionable. We propose a new experiment, joining laser fields
and heavy ions, which is feasible with present day technology and
which may indeed verify the theoretical prediction of adiabatic
pair creation. Our presentation relies on recent rigorous works in
mathematical physics.

\end{abstract}

\maketitle
\maketitle
\section{Introduction}

The creation of an electron positron pair in an almost stationary
 very strong external electromagnetic field (a potential
well) is often referred to as spontaneous pair creation
(\cite{greiner},\cite{alkofer},\cite{bamber}).
\begin{center}\label{pair}
\includegraphics[width=0.4
\textwidth]{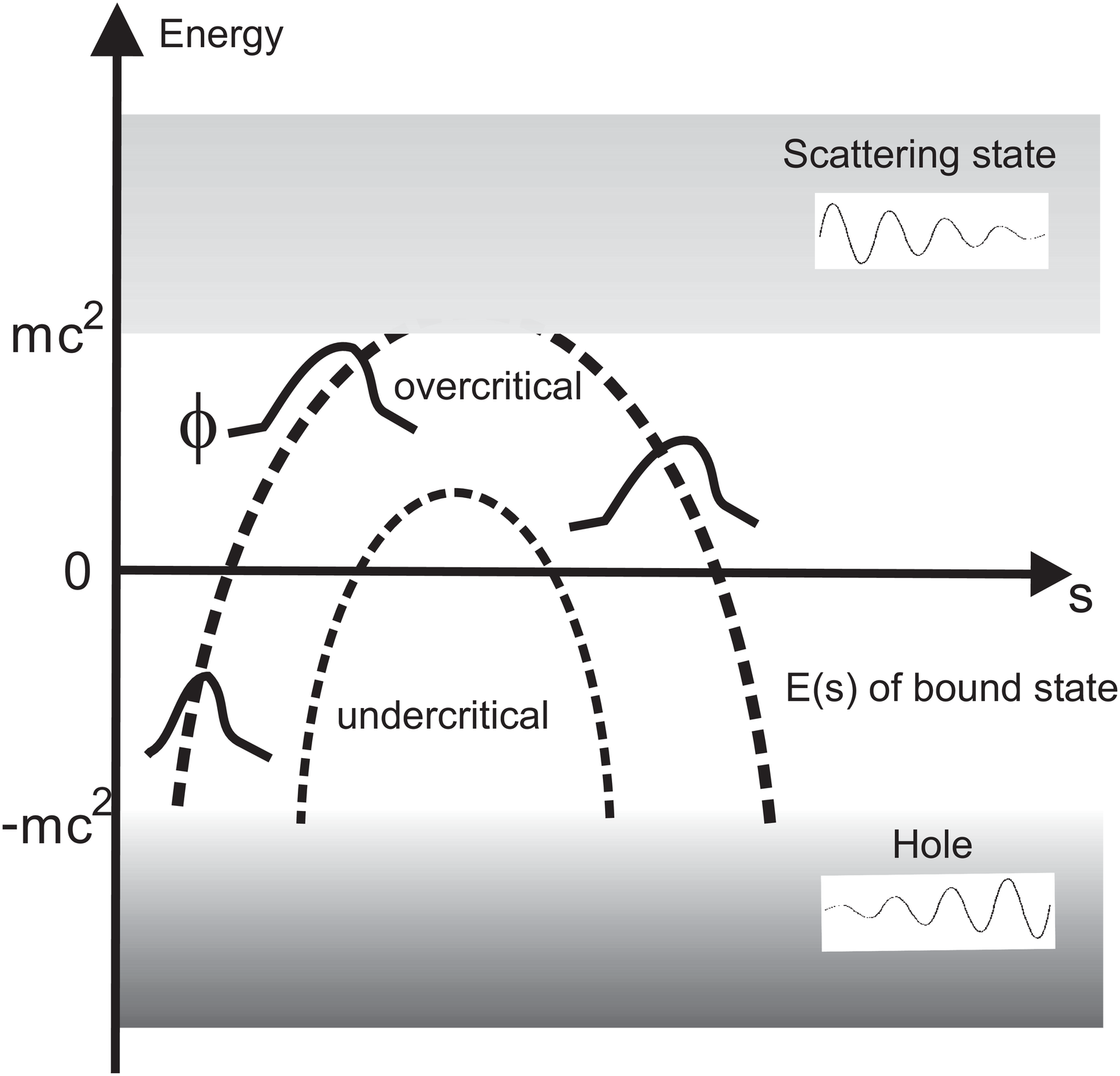}
\end{center}
This adiabatic
phenomenon emerges straightforwardly from the Dirac sea
interpretation of negative energy states: An adiabatically
increasing field (a potential well of changing deepness) lifts a
particle from the sea to the positive energy subspace (by the
adiabatic theorem) where it hopefully scatters and when the
potential is switched off one has one free electron and one
unoccupied state---a hole---in the sea \cite{beck}-\cite{gersh}
(see figure \ref{pair}). A better terminology---and the one we
shall use here---is thus adiabatic pair creation (APC).

The figure shows the bound state energy curve of a bound state
$\Phi$ emerging from the Dirac sea (vacuum), changing with time
$s$ due to the change of the potential well. If the potential
reaches a high enough value (critical value) the bound state
enters the upper continuum and scatters and if it has enough time
to escape from the range of the potential, before the potential
decreases under its critical value a pair will be created. The
experimental realization of APC has been extensively debated for
heavy ion collision. We note already here that the theoretical
descriptions concerning rates and line shapes were false. The main
reason is that no coherent theoretical approach had been found. We
give more details later.

Pair creation is also hotly debated in classical laser field technology nowadays
\cite{alkofer,bamber,ringwald,Blaschke,blaschke2}. But surprisingly the theoretical work on pair creation in
classical laser fields which is universally referred to as Spontaneous Pair Creation made no contact with
adiabatic theory which in fact it is. Indeed most of the theoretical work is  questionable and with that the
hope for experimental feasibility. We shall point out where present day arguments are dubious (reviewing the
relevant literature) and we shall present an argument and a new, well founded experimental possibility for pair
creation in a field which is a combination of an ion field and a classical laser field. We believe that this is
in fact the only possible experimental realization feasible in the near future. We shall give the details after
our theoretical treatment of APC which we do next.
%endchange

\section{Adiabatic Pair Creation}
 What we
 present here is the heuristic core of a mathematical proof of APC
(c.f.  \cite{picklneu,PDneu}). This proof while being
mathematically and technically  very involved is helpful to find
confidence in the heuristic short argument we shall give. The
mathematical proof changes the quantitative statements only
slightly.

Consider the one particle Dirac equation with external
electromagnetic field.
 On
microscopic time- and space-scales
\begin{equation}\label{tau}
\tau=\frac{mc^2}{\hbar}t=\frac{c}{\lambdabar_C}t\hspace{1cm}\mathbf{x}=\frac{mc}{\hbar}\mathbf{r}=\frac{\mathbf{r}}{\lambdabar_C}\end{equation}
the equation reads
\begin{eqnarray}\label{Diracmicalt}
   i\frac{\partial\psi_{\tau}}{\partial \tau}&=&-i\sum_{l=1}^{3}\alpha_{l}\partial_{l}\psi_{\tau}+A_{\varepsilon \tau}(\mathbf{x})\psi_{\tau}+\beta  \psi_{\tau}
   \nonumber\\&\equiv&(D^0+A_{\varepsilon
   \tau}(\mathbf{x}))\psi_{\tau}=D_{\varepsilon \tau}\psi_\tau
\end{eqnarray}
where $\varepsilon$ is the adiabatic parameter, representing the
slow time variation of the external potential and $A mc^2$ gives
the potential in the units $eV$ (we discuss later physical values
for $\varepsilon$).

We consider the Dirac equation on the macroscopic time scale
$s=\varepsilon\tau$:
\begin{eqnarray}\label{Diracmic}
   i\frac{\partial\psi_{s}}{\partial s}\equiv \frac{1}{\varepsilon}D_s\psi_{s}\;.
\end{eqnarray}
%Following \cite{nenciu} one can formulate the problem as follows: Do solutions $\psi_t$  of the Dirac equation
%exist with
%\begin{eqnarray}\label{loesung}
%&&\lim_{\varepsilon\rightarrow0}\lim_{s\rightarrow-\infty}\| P^- \psi_s\|=1
%%
%\\\nonumber\text{and }&&\lim_{\varepsilon\rightarrow0}\lim_{s\rightarrow+\infty}\| P^+ \psi_s\|=1\;?
%\end{eqnarray}
%
%($P^-$ and $P^+$ are the projectors of the negative (positive) part of the spectrum of the Dirac operator)
%change
 The
spectrum of the Dirac operator without external field is
$(-\infty,-1]\cap[1,\infty)$. The adiabatic theorem
ensures\cite{teufel} (for $\varepsilon$ small) that the gap can
only be crossed by bound states $\Phi_s$  of the Dirac operator
$D_s \Phi_s= E_s\Phi_s$ for which $E_s$ is a curve crossing the
gap (see figure). Let us call the curve a gap-bridge. {\it If
there is no gap-bridge the probability of pair creation is
exponentially small in $1/\varepsilon$  }. It is very important to
take note of this: No matter how steep the potential well is, i.e.
no matter how strong the ``force'' is, if there is no gap-bridge
pairs will only be created with exponentially small rate.

 We assume now, that
a gap-bridge exists and consider the bound state $\Phi_0$ (assumed
to be non degenerate) at the crossing.

We expand it in generalized eigenfunctions which of course depend
also on the "parameter" $s$. We shall need the eigenfunctions for
times $\sigma$ close to the critical time $\sigma=0$.

 Consider the eigenvalue equation
\begin{equation}\label{dgmp2}
D_{\sigma}\varphi= E \varphi
\end{equation}
for fixed $\sigma\in\mathbb{R}$.  The continuous subspace
%determined by $\mathbf{k}\in \mathbb{R}^3\backslash\{0\}$ or $\mathbf{k}=0$ and $\sigma\neq0$
is spanned by generalized eigenfunctions
$\varphi^{j}(\mathbf{k},\sigma,\mathbf{x})$, $j=1,2,3,4$, with
energy $E=\pm E_k=\pm\sqrt{k^2+1}$.
% $j=1,2$ being solutions with positive energy.
For ease of notation we
 will drop the spin index $j$ in what follows.

The generalized eigenfunctions also solve the Lippmann Schwinger
equation
\begin{eqnarray}\label{LS}
\lefteqn{\varphi
(\sigma,\mathbf{k},\mathbf{x})=\varphi_0(\mathbf{k},\mathbf{x})\nonumber}
\\&&+\int
 G^{+}_{k}(\mathbf{x}-\mathbf{x'}) A\hspace{-0.2cm}/_\sigma(\mathbf{x'})
 \varphi(\sigma,\mathbf{k},\mathbf{x'})d^{3}x'\;,
 \end{eqnarray}
with $\varphi_0(\mathbf{k},\mathbf{x})= \xi(\mathbf{k})
e^{i\,\mathbf{k}\cdot\mathbf{x}}$, the generalized eigenfunctions of
the free Dirac operator $D^0$, i.e.
  $G^{+}_{k}$ is the kernel of
$(E_{k}-D^0)^{-1}=\lim_{\delta\rightarrow0}(E_{k}-D^0+i\delta)^{-1}$
\cite{thaller}.

%\begin{equation}\label{kernel}
%G^{+}_{k}(\mathbf{x})=\frac{1}{4\pi}e^{ikx}\left(-x^{-1}(E_{k}+\sum_{j=1}^{3}\alpha_{j}k\frac{x_{j}}{x}+i\beta
%m) -ix^{-2}\sum_{j=1}^{3}\alpha_{j}\frac{x_{j}}{x}\right)\;.
%\end{equation}

Introducing the operator $T_\sigma^k$
\begin{equation}\label{tkdef}
T_\sigma^k f=\int
 G^{+}_{k}(\mathbf{x}-\mathbf{x'}) A_\sigma(\mathbf{x'})
 f(\mathbf{x'})d^{3}x',
 \end{equation}
(\ref{LS})  becomes
\begin{equation}\label{LSE}
(1-
T_\sigma^k)\varphi(\sigma,\mathbf{k},\cdot)=\varphi_0(\mathbf{k},\cdot)\;.
 \end{equation}
Note that
\begin{equation}\label{LSEbound}
(1- T_{0}^0)\Phi_0=0\;.
 \end{equation}

We  estimate the propagation of a wave function generated by the
{\em static} Dirac Operator $D_\sigma=D^0+A_\sigma(\mathbf{x})$,
where  $\sigma>0$ should be thought of as near the critical value
(the relevant regime turns out to be of order $\sigma=\mathcal{O}(
\varepsilon^{1/3})$).

Since the generalized eigenfunctions for
$(\sigma,\mathbf{k})\approx(0,0)$ are close to the bound state
$\Phi_0$ it is reasonable to write
 in leading
order:
\begin{equation}\label{dieistrichtig}
\varphi(\sigma,\mathbf{k},\mathbf{x})\approx\eta_\sigma(\mathbf{k})\Phi_0(\mathbf{x})\;.
\end{equation}
Since they solve (\ref{LS}), the first summand of (\ref{LS}) must
become negligible with respect to $\eta_\sigma(\mathbf{k})\Phi_0$,
which is part of the second summand. Hence
$\eta_\sigma(\mathbf{k})$ must diverge for $(\sigma,\mathbf{k})\to
(0,0)$.  For the outgoing asymptote of the state $\Phi_0$
(generalized Fourier transform) evolved with $D_\sigma$ near
criticality we have with (\ref{dieistrichtig}) that
\begin{eqnarray}\label{her}
\widehat{\Phi}_{out}(\sigma,\mathbf{k})&:=&\int
(2\pi)^{-\frac{3}{2}}
\Phi_0(\mathbf{x})\overline{\varphi}(\sigma,\mathbf{k},\mathbf{x})d^3
x\nonumber\\&\approx&(2\pi)^{-\frac{3}{2}}\overline{\eta}_\sigma(\mathbf{k})\;.
\end{eqnarray}
Now, for $(\sigma,k)$ close to but different from $(0,0)$,
$\eta_\sigma(\mathbf{k})\sim\overline{\widehat{\Phi}_{out}}(\sigma,\mathbf{k})$
will be peaked around a value $k(\sigma)$ with width
$\Delta(\sigma)$ (determined below) defined by
\begin{equation}\label{defdelta}
\eta_\sigma(k(\sigma)\pm\Delta(\sigma))\approx\eta_\sigma(k(\sigma))/\sqrt{2}\;.
\end{equation}
We may use the width for the rough estimate
\begin{equation}\label{ableitung}
|\partial_k\widehat{\Phi}_{out}(\sigma,\mathbf{k})|\  <
\Delta(\sigma)^{-1}\widehat{\Phi}_{out}(\sigma,k_\sigma)\;,
\end{equation}
where the right hand side should be multiplied by some appropriate
constant which we --- since it is not substantial --- take to be
one.
 Using
(\ref{dieistrichtig}), (\ref{her}), $d^3k=k^2d\Omega dk$ and
partial integration (observing
$\frac{\varepsilon}{-iks}\partial_ke^{-i(1+\frac{k^2}{2})\frac{s}{\varepsilon}}=e^{-i(1+\frac{k^2}{2})\frac{s}{\varepsilon}}$)
 we get

%$U_{\sigma}(s,0)\Phi_0=e^{-isD_{\sigma}}\Phi_0\approx e^{-i(1+\frac{k^2}{2})\frac{s}{\varepsilon}}\Phi_0$
\begin{eqnarray*}
U_{\sigma}(s,0)\Phi_0&=&e^{-isD_{\sigma}}\Phi_0\approx\frac{1}{(2\pi)^{\frac{3}{2}}}\int
e^{-i(1+\frac{k^2}{2})\frac{s}{\varepsilon}}\widehat{\Phi}_{out}\varphi
d^3 k
%\nonumber\\&=&\frac{1}{(2\pi)^{\frac{3}{2}}}\int\frac{m\varepsilon}{-iks}\partial_ke^{-i(m+\frac{k^2}{2m})\frac{s}{\varepsilon}}
%
%
%|\widehat{\Phi}_{out}|^2\varphi(x)k^24\pi d\Omega dk
%
\nonumber\\&=&\frac{-i\varepsilon}{s}
\int e^{-i(1+\frac{k^2}{2})\frac{s}{\varepsilon}}
%
%\nonumber\\&&
\partial_k\left(|\widehat{\Phi}_{out}|^2\Phi_0(x)k d\Omega\right) dk
\end{eqnarray*}
By (\ref{ableitung}), (\ref{her}) and (\ref{defdelta}), assuming
that $\Delta(\sigma)\ll k(\sigma)$
\begin{eqnarray*}
|\partial_k\left(|\widehat{\Phi}_{out}|^2k\right)|  d\Omega
dk&\approx& |\widehat{\Phi}_{out}|^2\left(\frac{2}{\Delta(\sigma)
k}+\frac{1}{k^2}\right) d^3k
\\&\approx&|\widehat{\Phi}_{out}|^2\frac{2}{\Delta(\sigma) k(\sigma)}d^3k
\;.
\end{eqnarray*}
Hence
\begin{eqnarray*}
|U_{\sigma}(s,0)\Phi_0(\mathbf{x})|&\leq&\frac{2\varepsilon|\Phi_0(x)|}{s\Delta(\sigma)
k(\sigma)}
\int
|\widehat{\Phi}_{out}|^2 d^3k\;.
\end{eqnarray*}
Since $\Phi_0$ is normalized we get for the decay time $s_d$,
defined by $|\langle U(s_d,0)\Phi_0,\Phi_0\rangle|\approx 1/2$
\begin{equation}\label{tautime}
s_d\approx 4\varepsilon (k(\sigma)\Delta(\sigma))^{-1}\;.
\end{equation}

The important information we must provide is thus
$\eta_\sigma(\mathbf{k})$ for $\sigma \approx 0 $. In view of
(\ref{LSE}) and (\ref{dieistrichtig}) we have that
\begin{equation}\label{LSE2}
(1-T_\sigma^k)\eta_\sigma(\mathbf{k})\Phi_0\approx\varphi_0(\mathbf{k},\cdot)\;.
 \end{equation}
We can estimate $\eta_\sigma(\mathbf{k})$ by considering the
scalar product of (\ref{LSE2}) with $A_0\Phi_0$:
\begin{eqnarray*}\label{LSE3}
\eta_\sigma(\mathbf{k})\langle(1-
T_\sigma^k)\Phi_0,A_0\Phi_0\rangle&\approx&\langle\varphi_0(\mathbf{k},\cdot),A_0\Phi_0\rangle\;.
 \end{eqnarray*}
One finds that
$\langle\varphi_0(\mathbf{k},\cdot),A_0\Phi_0\rangle\ =C k
+\mathcal{O}(k^2)$ with an appropriate $C\neq0$. Thus
\begin{eqnarray*}%\label{lemgeprobe2a}
\eta_\sigma(\mathbf{k})\approx Ck\langle
(1-T_\sigma^k)\Phi_0,A_0\Phi_0\rangle\rangle^{-1}\;.%\\&=&Ck\langle .
%(1-T_0^k)\Phi_0,A_\sigma\Phi_0\rangle\rangle^{-1}\; .
\end{eqnarray*}
 Expanding $T_\sigma^k$ in orders of $k$ around $k=0$  until fourth
order yields (the first order term turns out to be zero on general
grounds (\cite{picklneu},\cite{PDneu})
\begin{eqnarray*}\label{etareihe2}
\eta_\sigma(\mathbf{k})\approx
\frac{-Ck}{C_0\sigma-(|C_2|+\mathcal{O}(\sigma))
k^{2}-i(|C_3|+\mathcal{O}(\sigma))k^{3}}\;.
\end{eqnarray*}
For $C_0\sigma\approx C_2k^2$ the denominator behaves like $C_3
k^3$, otherwise it behaves like $C_0\sigma-C_2 k^{2}$. Hence by
(\ref{her})
\begin{eqnarray}\label{lemgeprobe}
|\widehat{\Phi}_{out}(\sigma,\mathbf{k})|^2\approx Ck^2\left(
(C_0\sigma-|C_2| k^{2})^2+|C_3|^2k^{6}\right)^{-1}
\end{eqnarray}
This result \cite{picklneu} differs from the results given in the
literature (see e.g. formula (7) in \cite{mprg}), we shall discuss
more details below. The right hand side of (\ref{lemgeprobe})
obviously diverges for $(\sigma,k)\rightarrow(0,0)$. For fixed
$0\neq\sigma\approx 0$ the divergent behavior is strongest close
to the resonance at ($C_0 \sigma- |C_2|k(\sigma)^2=0$)
\begin{equation}\label{knull}
k(\sigma)=\sqrt{\sigma C_0|C_2|^{-1}}
%=\mathcal{O}(\sqrt{\sigma})
\;.
\end{equation}
%At that point the right hand side of (\ref{lemgeprobe}) equals $-iC C_3^{-1}
%k(\sigma)^{-2}=\mathcal{O}(\sigma^{-1})$.
In view of (\ref{defdelta}) $\Delta(\sigma)$ can be roughly
estimated by setting the right hand side of (\ref{lemgeprobe})
equal to $1/2$ of its maximal size, i.e
$$C_0\sigma-|C_2| (k(\sigma)+\Delta(\sigma))^{2}\approx |C_3|k^{3}(\sigma)$$
hence
\begin{equation}\label{Delta}\Delta(\sigma)\approx
k(\sigma)^2|C_3| (2|C_2|)^{-1}
%=\mathcal{O}(\sigma)
\;.
\end{equation}

For a rough estimate of the decay time $s_d$ we set $\sigma= s_d$
and use (\ref{tautime}), (\ref{knull}) and (\ref{Delta}). This
yields in units of $\tau$ (cf. (\ref{tau}))
\begin{equation}\label{zeitskala}
s_d^{\frac{5}{2}}=
\frac{8\varepsilon|C_2|^{\frac{5}{2}}}{|C_0|^{\frac{3}{2}}|C_3|}\,.
\end{equation}

\section{Experiments}
\subsection{Heavy Ion Collisions} We turn now to the experimental
verifications and prior discussions in the theoretical physics
literature. The experimental verification of APC has been sought
in heavy ion collisions (HIC) (but without success so far
\cite{exp1}, \cite{exp2}). Here the adiabatic time scale on which
the field increases is directly determined by the relative speed
with which the heavy ions approach each other and one computes
that $\varepsilon$ is of order $10^{-1}$ \cite{greiner}.
Theoretical work on HIC was extensively done
\cite{greiner,mprg,PRLgreiner,mueller}. While the basic idea,
namely
 to find the shape of $\widehat{\Phi}_{out}$ (c.f.
\ref{lemgeprobe}), has been  seen before, the correct form and
meaning of $\widehat{\Phi}_{out}$ has not been arrived at.
Furthermore the analogue of our formula (\ref{lemgeprobe}) has
been interpreted as Breit Wigner shape, yet another source for
false rates.

Using (\ref{zeitskala}) $s_d({\rm
HIC})\propto\varepsilon^{\frac{2}{5}}$. The rigorous estimate
taking into account the time dependence of the external field,
yields $s_d\propto\varepsilon^{\frac{1}{3}}$ (\cite{PDneu}). Hence
if the field stays overcritical for times much larger than
$\varepsilon^{\frac{1}{3}}$ the probability of pair creation is
one, in the adiabatic case this is well satisfied. Thus HIC-APC is
theoretically proven. An interesting prediction is the shape of
the momentum distribution of the created positron. It would be
nice if the shape would be simply the resonance (\ref{lemgeprobe})
as suggested e.g. in \cite{mueller}. But that requires a somewhat
different situation than what one has in HIC. It would require an
overcritical static field (adiabatic is not enough) of a life time
much larger than the decay time $s_d$. Then in fact the resonance
(\ref{lemgeprobe}) would stay more or less intact (see also
\cite{reinhardt}). In an adiabatically changing field the
resonance (\ref{lemgeprobe}) changes however with the field and it
is highly unclear how.
\subsection{Laser-Pair-Creation} Another
experimental situation with adiabatically changing fields is
provided by lasers.  For laser fields (wavelength $\lambda$)
$\varepsilon=\lambda_C/\lambda\approx 10^{-6}$, where $\lambda_C$ is
the Compton wave length of the electron. It is in fact hoped, that
pair creation can be seen in a new generation of lasers which are
able to create in focus a very well localized overcritical classical
field. Unfortunately this hope is intermingled with misconceptions
which we shall try to sort out. Let us review shortly the present
status of the discussion on the possibility of creating pairs in
laser fields. An early computation by Schwinger \cite{schwinger}
predicted a pair creation rate for a constant strong electric field.
In that case the rate for small fields is exponentially small. This
is because pair creation arises in a similar manner as in the Klein
paradox tunnelling, where ``negative energy states'' overlap with
their exponentially small tails the ``positive energy states''
\cite{kleinT}. One can picture the situation by ``tilting'' the
spectral gap (of figure \ref{pair}) under the influence of the
electric field) so that wave functions ``can reach across''.  As a
rule of thumb the critical electric field $E_c$ which one needs is
estimated by the potential energy the electron acquires over a
distance of roughly the Compton-wave length:
$eE_c\lambdabar_C=2mc^2$. The Schwinger computation is certainly
correct but of course  has nothing to do with lasers.
%change
The Schwinger-computation has been generalized by Brezin and Itzykson  to time varying but spatially constant
electric fields \cite{itzyk}. That generalization did not change the rates for pair creation as computed by
Schwinger. They are still  determined by Klein-paradox tunnelling across the ``tilted'' spectral gap. But Brezin
and Itzykson apply now their results to laser fields. Their argument is, that while laser fields are not of the
type of the electric potential field they treat (a field of a huge capacitor with plates infinitely far apart)
since they have magnetic parts as well, their field strength should  be at least a lower bound to the strength
of a realistic laser field which would be able to create pairs. That the magnetic part of the laser field can be
neglected is not much argued for, except that it is bluntly stated that pure magnetic fields cannot create
pairs. In \cite{itzykzuber} one finds a perturbation argument from which it is also concluded that pair creation
is an ``electric effect''. As a note aside it is also folklore wisdom  that a single electro magnetic wave
cannot produce pairs, it is therefore standardly assumed that at least two laser fields are superposed so that
in the focal region a standing wave is formed. The Brezin-Itzykson reduction of a laser field to an electric
potential field became the basis of all further theoretical research on laser pair creation (see e.g
\cite{alkofer,bamber,ringwald}).

What is wrong with all that? Firstly we recall that
 the Brezin-Itzykson treatment was intended to discourage hopes
 for seeing pair creation with optical lasers. They actually say
 that they would not have trusted their bound if it would have turned out
 lower than what they got. The moral which should be drawn is simply: The
 true lower bound can and will  be much higher than what Brezin-Itzykson found.
 In
 fact we shall argue later that no lower bound exists. In reverse,
 it is
not a priori clear at all that pair creation is determined by the
maximal value of the present $\vec E$-field.  What is clear
however is that that treatment gives wrong results when applied to
a single overcritical laser beam, where no pair creation occurs.
Therefore if the Brezin-Itzykson treatment is thought to be of
relevance for superposed laser fields an argument is truly needed,
but none has been given.

Secondly we wish to emphasize the obvious, namely that the talk
about electric fields in the context of quantum mechanics and in
particular in the context of  pair creation can be misleading:
Only the four-vector-potential enters in the Dirac equation.
Descriptions in terms of the {\em local} behavior of the $E-$ or
$B-$ fields may therefore lead off the track. (This is similar to
the role of the $A$ and $B$ field in the Aharonov Bohm effect). In
particular laser fields must not  be modelled by a constant
electric field. The $A$-fields with $A_0=0$ (in Coulomb gauge)
differs  from the $A$-field where only $A_0\ne0$ (in Coulomb
gauge) in all respects. In particular it is important to take the
full geometry of the laser field into account.  One moral of APC
is that pair creation is a ``global'' effect, i.e. the global
$\mathbf{x}$ and $t$ dependence of the field plays a role:
Criticality has to be defined via the bound states of the {\it
potential}. Weather there is a bound state at the edge or not
depends on the full $\mathbf{x}$-dependence on that field, not on
the maximal $\vec E$-field only. In APC the pair creation behavior
of an (slightly) overcritical external potential is qualitatively
different from the pair creation behavior of an (slightly)
undercritical potential, though their ``local'' $\vec E$- and
$\vec B$-fields (in particular their maximal $\vec E$-field) may
be almost the same. That clearly contradicts the assumption that
the pair creation rate is given by the maximal $\vec E$-field
only.  A related point arising from the global nature of APC is
that perturbative treatments are highly questionable \cite{popov}.

Let us now return to the crucial question: Does there exist a
gap-bridge for laser fields? The electric part of a laser field is
not a gradient of an electric potential and unfit for the job to
lift a ground state from the lower to the upper continuum. What
about taking the laser field seriously, considering also the
magnetic part of the field? The magnetic field, in particular the
{\em spin-magnetic field interaction} provides in fact an
adiabatically changing potential as APC requires. Unfortunately that
does not work either. For constant $\vec B$- fields it is already
known that no gap-bridge exists (see \cite{achu} and \cite{loss} for
a rigorous statement).  We give the simple argument now for the more
general case of fields with varying magnitude but fixed direction.
The argument shows that typically a gap-bridge does not exist for
the Dirac equation
 {\it without electric potential ($A_0=0$ in Coulomb gauge)},% and {\it localized} $\vec A$-field
 i.e. that they do not exist
  for laser fields. %(our argument does not rule out $\vec A$-fields which do not tend to zero as $x\to\infty$. Such fields may allow gap-bridges \cite{achu}, but seem somewhat academical  and do not fit for laser fields).
  In fact we argue that there is no bridge crossing the spectral value zero!  Starting with the eigenvalue
equation for the Dirac operator and writing $\underline{A}=\vec{A}\cdot\vec{\alpha}=\sum_{j=1}^3A_j\alpha_j$ we
compute
\begin{eqnarray*}
(D_0+\underline{A})\Phi&=&(E-A_0)\Phi\\
(D_0+\underline{A})^2\Phi&=&(D_0+\underline{A})(E-A_0)\Phi\\
((i\underline{\nabla}-\underline{A})^2+1) \Phi &=&
(i(\underline{\nabla} A_0) + (E-A_0)^2)\Phi\,.
\end{eqnarray*}
This then yields
\begin{eqnarray*}
(H+m^2+i\underline{\nabla}\, \underline{A}) \Phi &=&
(i(\underline{\nabla} A_0) + (E-A_0)^2)\Phi
\end{eqnarray*}
where $H$ is the Schr\"odinger operator with vector potential
$\vec{A}=(A_1,A_2,A_3)$ in Coulomb gauge (${\rm div} \vec{A}=0$).
By simple $\alpha$-algebra we have that
$$\underline{\nabla}\;\underline{A}=\sum_{j,k}\alpha_j\alpha_k\partial_j
A_k={\rm div} \vec{A} \mathbb{E}_4+\underline{B}
\alpha_4=\underline{B} \alpha_4$$ where $\mathbb{E}_4$ is the
$4\times 4$-unit matrix,
$$\alpha_4=\alpha_1\alpha_2\alpha_3=\left(%
\begin{array}{cc}
  0 & \mathbb{E}_2 \\
  \mathbb{E}_2 & 0 \\
\end{array}%
\right)$$ and $\underline{B}=\vec{B}\cdot\vec{\alpha}$. Now we
assume that the magnetic field $\vec{B}$ has constant direction
(as it is the case in a standing wave), i.e.
$\underline{B}(\vec{x}) \alpha_4=b(\vec x)
\underline{B}_0\alpha_4$ with constant matrix $\underline{B}_0$
and real function $b(\vec x)$ giving the strength of the magnetic
field.

Using this and $A_0=0$ we have that
\begin{eqnarray*}
(H+1+ b(\vec x) \underline{B}_0\alpha_4)\Phi &=& E^2\Phi
\end{eqnarray*}
%Assuming $\underline{\nabla}\, \underline{A}$ being constant (constant magnetic field $\mathbf{B}$)
We go to the eigenspaces  of the matrix $\underline{B}_0\alpha_4$
setting $\varphi_j=\langle s_j,\Phi\rangle$. Note here, that
$\alpha_4$ commutes with $\alpha_l$ for $l=1,2,3$ and that the
eigenvalues of the matrix $\underline{B}_0\alpha_4$ are $\mu=\pm
1$
\begin{eqnarray}\label{schroe}
(H+1+\mu b(\vec x) ) \varphi_j&=& E^2\varphi_j\;.
\end{eqnarray}
(Note aside that in physical units $\mu$ is the magnetic moment.)
From this we see that the energy curve of the Schr\"odinger operator
$H+1+\mu b(\vec x)$ must be quadratically zero if there is a gap
bridge of the corresponding Dirac operator crossing zero. This
however is a very special behavior of eigenvalues and thus unlikely.
(See \cite{achu} where a famous special case is actually discussed.)
One may conclude that gap bridges do not occur at all for laser
fields. On the other hand if $A_0\neq 0$ this argument breaks down
and a gap bridge may exist.

Our first conclusion is thus: No matter how tricky laser waves are superposed and how strong the fields may be,
no gap-bridge arises and  pair creation will at most be exponentially small in $1/\varepsilon$. Note however,
that we made an external field approximation in our treatment. Considering self interaction effects, in
particular the anomalous magnetic moment of the electron, a gap bridge may arise for $\vec B$ fields which are
more than a few thousand times overcritical \cite{neutronen}.

Our second conclusion is however: Since the magnetic field in
junction with a Coulomb field can produce a gap bridge, a new (and
possibly the only one using lasers) experimental possibility for
APC arises. We propose to combine laser- and heavy ion fields,
i.e. to shoot heavy ions into the focus of the laser. We model the
laser field by an oscillating magnetic field $\vec B$ constant in
space. First let us explain, why this model is satisfied for the
present situation. As mentioned above, descriptions in terms of
$\vec E$ and $\vec B$ fields may be misleading, so let us model
the vector potential of the laser. In Coulomb-gauge the vector
potential has no $A_0$-component. Since the vector potential of an
optical laser varies for a given time only slightly over the range
of the heavy ion potential the laser-field is in good
approximation given by its first order Taylor expansion around the
center of the nucleus. Note, that in Coulomb-gauge the zero-order
term of the vector potential is zero. Hence the leading order term
of the vector potential is the first order term, which has (in
Coulomb gauge) no gradient and can thus be written in the form
$\vec{B}\times\vec{x}$ for some $\vec{B}$ constant in space,
oscillating in time. Thus the vector potential of the laser is in
good approximation the potential of a spatially constant magnetic
field in the situation at hand.

In \cite{loss} an estimate (actually a lower and upper bound) for
the  the critical strength of a (spatially constant) magnetic
field which has a gap bridge in presence of a nucleus with charge
$Z$ is given. Instead of estimating the lower bound on the $\vec
B$-field needed in junction with the ion field to have APC we
translate the corresponding lower bounds to the laser electric
field strength $E_<$. This enables us to have easy comparison with
the results given by the laser community. We should however warn
the reader that we use here the  the lower bound of \cite{loss}
because it is computable number,  while the (reliable) upper bound
is not sharp defined.

\begin{table}

\begin{tabular}{|l|r|r|}
\hline
  Ion & Z & $E_{<}/E_c$ \\
  \hline\hline
  H & 1 & $\approx 15.000$ \\
  \hline
  C & 12 & $\approx 100$ \\
  \hline
  U & 92 & $\approx 2$ \\
  \hline
\end{tabular}
\caption{\label{tabelle} The table gives an estimated lower bound
on the electric field strength of a laser for a gap bridge, when a
nucleus with charge $Z$ is in the focus of the laser. The lower
bound $E_{<}$ is given in units of the critical Schwinger field
strength $E_c$.}
\end{table}

The lower bound is
$E_<\approx\frac{4}{5(Z\alpha)^2}\frac{m^2c^2}{e\hbar}$, i.e. only
if the electric field strength $E_{APC}$  fulfills $E_{\rm
APC}>E_<$ one can expect that APC takes place. In table
\ref{tabelle} we list the values for $E_<$ for three different
nuclei in units $E_c$. We recall that $E_c=\frac{m^2c^2}{e\hbar}$
(the critical field strength given by Schwinger) is the critical
field strength where the respective $B$-field in a light wave
satisfies $\mu B=mc^2$. For uranium this yields that the field has
to be two times overcritical (see table \ref{tabelle}). Laser
fields which are a few times overcritical are expected to be
reached by a new generation of lasers, called XFEL (see e.g.
\cite{ringwald}). Bringing a heavy ion (like uranium) into the
focus of such a laser APC should be observable. In fact, the
transition from no pairs to periodic (twice per laser period)
appearance of pairs when the laser field becomes strong enough
(twice overcritical, according to table \ref{tabelle}) is very
much like a phase transition.

{\bf Acknowledgement}: We thank Herbert Spohn for triggering the
application to lasers. We thank  ESI (Vienna) for hospitality and
funds. Work was partly funded by DFG.

\end{document}